\documentstyle[12pt]{article}
\begin{document}

\begin{center}
{\Large {\bf Stochastically perturbed flows: Delayed and interrupted 
evolution}}\\
V. Balakrishnan*$\footnote{Permanent address: Department of Physics, 
Indian Institute of Technology-Madras, Chennai 600 036, India}$, 
C. Van den Broeck,** and I. Bena**\\
{\it *Centre for Nonlinear Phenomena and Complex Systems, 
Universit\'e Libre de Bruxelles, C.P. 231, Bd. du Triomphe, 1050 
Brussels, Belgium} \\ 
{\it **Limburgs Universitair Centrum, 
B-3590 Diepenbeek, Belgium}
\end{center}

\vspace{2cm}
{\bf Abstract}\\
We present analytical expressions for the time-dependent and 
stationary probability distributions  
corresponding to a stochastically perturbed one-dimensional flow
with critical points, in two physically relevant situations: 
{\it delayed evolution}, in
which the flow alternates with a quiescent state in which the 
variate remains frozen at 
its current value for random intervals of time; and 
{\it interrupted evolution}, in which 
the variate is also re-set in the quiescent state to 
a random value drawn from a fixed distribution. In the former case,
the effect of the 
delay upon the first passage time statistics is analyzed. In the 
latter case, the conditions under which an extended stationary 
distribution can exist as a consequence of the competition between
an attractor in 
the flow and the random re-setting are examined. We elucidate
the role of the normalization 
condition 
in eliminating the 
singularities arising from the unstable critical points of the flow, 
and present a  
number of representative examples. A simple formula is obtained
for the stationary distribution
and interpreted physically. A similar interpretation is also given 
for the known formula for the stationary 
distribution in a full-fledged dichotomous flow.\\ 


\newpage

\section{Introduction}
Physical systems are invariably subject to
stochastic perturbations of various kinds. 
The effects of noise upon the evolution of classical 
dynamical systems have been studied quite extensively for several 
decades now \cite{horst}-
\cite{garcia}. While much of the 
work in the area has been concerned with Gaussian noise (which  
offers several analytical advantages), other forms of 
noise are also relevant in specific instances.
The inclusion of a finite correlation time for the noise 
(a closer approximation to physical reality) has  also  
led to the uncovering of several interesting features that are absent 
in the idealized case of delta-correlated noise. 

In many physical situations, the stochasticity affects the dynamics
more strongly than as just
a small perturbation of the deterministic evolution. An important
case is that of random interruption of the  
deterministic flow of a system variable $x$;
during each
interruption (the `quiescent' state), $x$ either remains frozen
at its current 
value, or is re-set to a random value drawn from a 
prescribed distribution, and remains frozen there.
Further evolution (the `active state')
is resumed after a random 
interval of time. 
We call these possibilities {\it delayed evolution} 
and {\it interrupted evolution}, respectively. 
A main purpose of this 
paper is to obtain exact expressions for both the
time-dependent and
the stationary probability distributions of the 
driven variable $x$ in both these situations.
While problems of this general type 
have been studied right from the early days  
in a variety of specific contexts ranging from 
chromatography \cite{giddings} to the collision broadening
of line shapes \cite{rautian} and other relaxation phenomena
\cite{datta}, our focus here is on an altogether different
aspect: the elucidation of the role of the stable and unstable
critical points of the flow in determining these distributions. 

The random interruptions are most simply modeled by assuming
that they are switched on and off according to a
Poisson pulse process of intensity $\lambda$ \cite{poisson}. The mean
duration
$\lambda^{-1}$ of the active and quiescent states provides a 
time scale whose interplay with the characteristic time scale
of the deterministic evolution will be of interest. (For simplicity
we assume that the switching between the two states occurs at
a common mean rate $\lambda$. All our results can be easily extended 
to the case of unequal switching rates  
in a straightforward manner.) We identify 
the active and quiescent 
states by the state labels $\xi = +1$ and
$\xi = -1$ respectively. It is useful to note that the random
variable $\xi(t)$ is in fact a  symmetric, stationary dichotomous Markov 
process (DMP) with a correlation time $ \tau_{c} =(2\lambda)^{-1}$. The 
conditional probability density $P(x,\xi,t\vert x_{0}, 
\xi_{0})$ will occasionally be abbreviated to $P_{\pm}(x,t)$ when it is 
convenient to do so without causing confusion.

\section{Delayed evolution}
\subsection{Solution for  $P(x,\xi,t\vert x_{0},\xi_{0})$} 
Let the deterministic evolution (in the active or
$\xi = +1$ state) be given by 
\begin{equation}
\dot{x} = f_{+}(x)\, .
\label{fplus}
\end{equation}
The first step is to identify the critical points of the flow given 
by Eq. (\ref{fplus}), i.e., the roots of $f_{+}(x) = 0$, and the 
direction in which the flow occurs in each interval between successive 
critical points. Clearly, for 
an initial condition $x(0) = x_{0}$ lying between two such points, 
$x(t)$ is restricted to this interval for all $t \geq 0$. We 
therefore expect $P_{\pm}(x,t)$ to involve step functions that 
reflect this fact. 

The device we use to solve the problem of obtaining the exact 
time-dependent probability distributions for delayed evolution is to 
map the problem onto that of a particular dichotomous flow.
Since $\dot{x}$ vanishes identically in the state $\xi = -1$, 
the evolution equation for  $x$ at {\it any} time $t$ 
can be written as
\begin{equation}
\dot{x} = \left(\frac {1+\xi (t)}{2}\right)f_{+}(x)\, .
\label{evoleq}
\end{equation}
Equation (\ref{evoleq}) implies that $\{x(t),\xi(t)\}$ constitute a 
stationary Markov process. 
Denoting the anti-derivative of $1/f_{+}(x)$ by $q(x)$, i.e., 
$q(x) = \int dx/f_{+}(x)$, we see that $\dot{q}$ is a 
DMP that switches between the values 1 and 0. Therefore the 
(non-stationary!) process 
$y = q - \frac{1}{2}t$ is given by the stochastic differential 
equation $\dot{y} = \zeta (t)$ where $\zeta (t)$ is a DMP switching 
between the values $\frac{1}{2}$ and $-\frac{1}{2}$: in other words, 
$y$ is a persistent diffusion process \cite{bc}. Its conditional densities 
$P_{\pm}(y,t)$ (retaining the symbol $P$ for these, in a slight abuse 
of notation) therefore satisfy the telegrapher's equation
\begin{equation}
\left(\frac{\partial^{2}}{\partial t^{2}} + 
2\lambda \frac{\partial}{\partial t}  - 
\frac{1}{4} \frac{\partial^{2}}{\partial y^{2}} 
\right)P_{\pm}(y,t) = 0 \,.
\label{teleg}
\end{equation}
The general solution of Eq. (\ref{teleg}) can be written down
\cite{koshly} (using, for instance, the method of characteristics), 
once the initial conditions $P_{\pm}(y,0)$ 
and $\dot{P}_{\pm}(y,0)$ are specified. To obtain the latter, we
begin with the initial conditions on the original densities, namely,
\begin{equation}
P(x,\pm,0\vert x_{0},\pm) = \delta (x-x_{0})\,\,,\,\, P(x,\pm,0\vert 
x_{0},\mp) = 0\,\,.
\label{icx}
\end{equation}
These translate into the conditions 
\begin{equation}
P(y,\pm,0\vert y_{0},\pm) = \delta (y-y_{0})\,\,,\,\, P(y,\pm,0\vert 
y_{0},\mp) = 0
\label{icy}
\end{equation}
for the densities of the $y$-process, where $y_{0} \equiv q(x_{0})$. The 
initial conditions on the time derivatives of the densities are 
found by recalling that $P_{+}(y,t)$ and $P_{-}(y,t)$ actually satisfy 
the coupled first order equations 
$(\partial_{t} \pm  \frac{1}{2}\partial_{y} 
+\lambda)P_{\pm}(y,t)= \lambda P_{\mp}(y,t)$. Insertion of Eqs. 
(\ref{icy}) in the latter leads to the 
initial conditions 
\begin{equation}
\dot{P}(y,\pm,0\vert y_{0},\pm) = \mp\frac{1}{2}\delta\,' (y-y_{0}) 
- \lambda \delta (y-y_{0})\,\,,\,\,
\dot{P}(y,\pm,0\vert y_{0},\mp) =\lambda \delta (y-y_{0})\,\,. 
\label{icydot}
\end{equation}
Using Eqs.(\ref{icy}) and (\ref{icydot}) in the general solution of 
Eq. (\ref{teleg}) and simplifying, we finally arrive at the  
solutions given in Eqs. (\ref{plusplus})-(\ref{minusminus}) below
for the set of densities $P(x,\xi,t\vert x_{0},\xi_{0})$. 
Let  
\begin{equation}
T(x_{0}, x) \equiv \int_{x_{0}}^{x} dx'/f_{+}(x') = q(x) - q(x_{0})
\label{T}
\end{equation}
denote the time taken to move from $x_{0}$ to a given point $x$ under the 
(autonomous) flow of Eq. 
(\ref{fplus}). Correspondingly, the location 
of the point  reached under this flow in a given
time interval $t$, starting from $x_{0}$, is given by the expression  
\begin{equation}
q^{-1}(q(x_{0}) +t) \equiv X(x_{0},t)\,\,.
\label{X}
\end{equation}
In reverting from $y$ to the original variable $x$, use is made of the
relationship 
\begin{equation}
\delta \left(x-X(x_{0},t)\right) =
\frac{\delta \left(t-T(x_{0}, x)\right)}{ \vert f_{+}(x) \vert}\,\,.
\label{deltaXT} 
\end{equation} 
Then, omitting the arguments in $X(x_{0},t)$ and 
$T(x_{0}, x)$ for brevity, we find:  
\begin{eqnarray}
P(x,+,t\vert x_{0}, +) &=&  e^{-\lambda t}\,\bigg[ \delta (x- X) 
+{} \nonumber \\ 
& +&{} 
\frac{\lambda}{\vert f_{+}(x)\vert}\left(\frac{T}{t - 
T}\right)^{\frac{1}{2}}\,I_{1}\left(2\lambda \sqrt{T(t - 
T)}\right)\,\theta (x-x_{0})\,\theta (X -x)\bigg]\,\,,
\label{plusplus}    
\end{eqnarray}
\begin{eqnarray}
P(x,+,t\vert x_{0}, -) &=& P(x,-,t\vert x_{0}, +)\nonumber \\
&=& \frac{\lambda e^{-\lambda t}}{\vert f_{+}(x)\vert}\,
I_{0}\left(2\lambda \sqrt{T(t - 
T)}\right)\,\theta (x-x_{0})\,\theta (X -x)\,\,,
\label{plusminus}    
\end{eqnarray}
and
\begin{eqnarray}
P(x,-,t\vert x_{0}, -) &=&  e^{-\lambda t}\,\bigg[ \delta (x- x_{0}) 
+{} \nonumber \\ 
& +&{} 
\frac{\lambda}{\vert f_{+}(x) \vert}\left(\frac{T-t}{T}\right)^{\frac{1}{2}}\,
I_{1}\left(2\lambda \sqrt{T(t - 
T)}\right)\,\theta (x-x_{0})\,\theta (X -x)\bigg]\,\,,
\label{minusminus}    
\end{eqnarray}
where $I_{n}$ is the modified Bessel function of order $n$. The
step functions in Eqs. (\ref{plusplus})-(\ref{minusminus})  refer to 
the situation in which $f_{+}(x) > 0$ in the interval concerned. They
originally appear as 
$\theta(T)\,\theta(t-T)$, so that when $f_{+}(x) < 0$ 
the product of step functions in  these equations  is replaced by 
$\theta(x_{0}-x)\,\theta(x-X)$.

\subsection{First passage time distributions}

An interesting question that arises naturally in the case of evolution 
that is stochastically delayed as above is the following: what is the
effect of the 
delay on the corresponding level crossing (or first passage time) 
statistics? As the flow does not reverse direction in any interval 
between two successive critical points, it is 
clear that $x (t)$ is a non-decreasing [respectively, 
non-increasing] function of $t$
for all $t\geq 0$ if $f_{+}(x) > 0$ [respectively, $f_{+}(x) < 0$] in 
the interval concerned. It follows at once that
$P(x,+,t\vert x_{0},\xi_{0})\, dx $ is {\it also} equal to the probability  
$Q(t,x\vert x_{0},\xi_{0}) \,dt $ of reaching the point  $x$ for 
the first time at 
time $t$. Thus the quantity 
$\vert f_{+}(x) \vert \,P(x,+,t\vert x_{0},\xi_{0}) \equiv 
Q(t,x\vert x_{0},\xi_{0})$, 
expressed in terms of the appropriate variable (namely, $t$), 
is the first 
passage time density for reaching the point $x$, starting 
at $t=0$ from the 
point $x_{0}$ and in the state $\xi_{0}$. With the help of the useful 
relationship in Eq. (\ref{deltaXT}) we obtain, after a bit of 
simplification, 
\begin{equation}
Q(t,x \vert x_{0}, +) =  e^{-\lambda t}\,\bigg[ \delta (t- T) 
+\frac{d}{dt}
I_{0}\left(2\lambda \sqrt{T(t - 
T)}\right)\,\theta (t-T)\bigg]\,\,
\label{Qplus}    
\end{equation}
and
\begin{equation}
Q(t,x \vert x_{0}, -) =  \lambda e^{-\lambda t}
I_{0}\left(2\lambda \sqrt{T(t - 
T)}\right)\,\theta (t-T)\,\,.
\label{Qminus}    
\end{equation}
The corresponding generating functions (i.e., Laplace transforms) 
are given by relatively simple expressions: 
 
\begin{equation}
\tilde{Q}(s,x \vert x_{0}, +) =  e^{-(s+\lambda) T}\,
e^{\lambda^{2}T/(s+\lambda)}
\label{Qplustilde}    
\end{equation}
and
\begin{equation}
\tilde{Q}(s,x \vert x_{0}, -) = \frac{\lambda}{(s+\lambda)}
e^{-(s+\lambda) T}\,
e^{\lambda^{2}T/(s+\lambda)}\,.
\label{Qminustilde}    
\end{equation}
As these expressions reduce to unity when $s=0$,  it is evident 
that the first passage time distributions concerned are properly
normalized.

The mean first passage time is then  
\begin{equation}
\langle t(x_{0}\rightarrow x)\rangle_{+} = 2\,T(x_{0},x)\,\,\,,\,\,\,
\langle t(x_{0}\rightarrow x)\rangle_{-} = 2\,T(x_{0},x)+ 
\lambda^{-1}\,,
\label{meanfptpm}    
\end{equation}
where the subscripts on the angular brackets indicate the starting 
state. The extra  $\lambda^{-1}$ in 
$ \langle t \rangle_{-}$ is understood as follows. We expect  
$ \langle t \rangle_{-}$ to exceed $ \langle t \rangle_{+}$ 
by just the
mean forward recurrence time in the $\xi = -1$ state; and since the
DMP is governed by an uncorrelated Poisson process 
(implying an {\it exponential} 
waiting time density), this is the same as the 
mean duration of this state, i.e., $\lambda^{-1}$.

The mean squares are 
\begin{equation}
\langle t^{2}(x_{0}\rightarrow x)\rangle_{+} = 4\,T^{2} 
+ 2\,T \lambda^{-1}\,\,\,,\,\,\,
\langle t^{2}(x_{0}\rightarrow x)\rangle_{-} = 4\,T^{2} + 6\,T \lambda^{-1}
+ 2\,\lambda^{-2}\,.
\label{meansqfptpm}    
\end{equation}
The stationary {\it a priori} probabilities of being in 
the active and quiescent states 
being equal, the net mean first passage time  and its mean square are 
therefore 
\begin{equation}
\langle t(x_{0}\rightarrow x)\rangle = \frac{1}{2} 
(\langle t\rangle_{+} +
\langle t\rangle_{-}) =
2\,T(x_{0},x) + 
(2\lambda)^{-1} = 2\,T(x_{0},x) + \tau_{c}
\label{meanfpt}    
\end{equation}
and
\begin{equation}
\langle t^{2}(x_{0}\rightarrow x)\rangle = \frac{1}{2} 
(\langle t^{2}\rangle_{+} +
\langle t^{2}\rangle_{-}) =
(2\,T(x_{0},x) + 
\lambda^{-1})^{2} = 4\,(T(x_{0},x) + \tau_{c})^{2}\,\,,
\label{meansqfpt}    
\end{equation}
where we recall that $\tau_{c} = (2\lambda)^{-1}$ is the correlation 
time of the DMP $\xi(t)$ controlling 
the switching between the evolving and quiescent states. The relative 
fluctuation in the first passage time (the ratio of its standard 
deviation to  its mean) is therefore given by
\begin{equation}
\Delta t(x_{0}\rightarrow x) =  
\frac{\sqrt{\tau_{c}(4T+3\tau_{c})}}{2\,T +\tau_{c}}\,\,.
\label{relfln}    
\end{equation}
We note that $\Delta t$ tends to a constant 
value ($=\surd 3$) as $T(x_{0},x)/\tau_{c} \rightarrow 0$, while it 
decays like $(T(x_{0},x)/\tau_{c})^{-\frac{1}{2}}$ for very large 
values of this ratio. 

These results can easily be generalized to the case of unequal mean durations 
of the evolving and quiescent states.

\section{Interrupted evolution}
\subsection{Solution for $P(x,\xi,t\vert x_{0},\xi_{0})$}

We turn now to the case when $x$ is {\it re-set} instantaneously 
to a random value 
drawn from a prescribed normalized distribution $\phi_{-}(x)$ 
whenever a transition occurs from the active or evolving state to the 
quiescent state, and remains fixed at that value till the active state 
occurs again. It is evident that the 
re-setting can be taken to occur at any instant during the
quiescent state, without affecting the results.
We note that $x$ may be re-set randomly anywhere in 
its total range $(-\infty, \infty)$, and not just in the interval between the 
two successive zeroes of $f_{+}(x)$ that contains its initial value $x_{0}$. 

Once again, it is helpful to regard the switching between states 
as being triggered by the DMP $\xi (t)$. Denoting the transition 
probabilities for this DMP by  $p(\xi,t\vert \xi_{0})$, we have 
\begin{equation}
p(\pm, t \vert \pm) = e^{-\lambda t} \cosh \lambda t\,\,,\,\,
p(\pm, t \vert \mp) = e^{-\lambda t} \sinh \lambda t\,\,.    
\label{p}    
\end{equation}
Each of these probabilities tends to $\frac{1}{2}$ as $t \rightarrow 
\infty$. We therefore have the limit
\begin{equation}
\lim_{t \rightarrow \infty} P(x,-,t\vert x_{0},\xi_{0}) = 
\frac{1}{2}\,\phi_{-}(x)\,\,,
\label{Pminuslimit}
\end{equation}
where
\begin{equation}
\int_{-\infty}^{\infty} \phi_{-}(x) dx = 1 \,\,.   
\label{phinorm}
\end{equation}
Similarly, we may ask whether 
\begin{equation}
\lim_{t \rightarrow \infty} P(x,+,t\vert x_{0},\xi_{0}) = 
\frac{1}{2}\,\phi_{+}(x)\,\,,
\label{Ppluslimit}
\end{equation}
where $\phi_{+}(x)$ represents the stationary probability density
of $x$ in the 
active state. The conditions under  which 
a normalizable, extended density  $\phi_{+}(x)$ exists
will be examined in detail in the next 
subsection.

Owing to the instantaneous re-setting of $x$ 
in each occurrence of 
the quiescent state, it is no longer useful to map the problem
to a stochastic differential equation. However, it is
clear that it is only the {\it last} such 
re-setting that controls the location of the final point 
$x$ at time $t$, in each realization of the random process.  
The exact expressions for  $P(x,\xi,t\vert x_{0},\xi_{0})$ are
therefore easily found by enumerating 
the processes contributing to the probabilities 
and summing over the corresponding 
propagators. As  $\xi (t)$ is governed by a 
Poisson process of intensity $\lambda$, the 
probability of a flip in $\xi (t)$ occurring in an infinitesimal 
interval $\delta t$ is $\lambda \delta t$, while the probability of 
the persistence of either state of $\xi$ for an interval $t$ is 
$\exp (-\lambda t)$. 
Using these facts, and starting with the simplest 
case, we find
\begin{eqnarray}
P(x,-,t\vert x_{0},+)& = &\sum_{n=0}^{\infty} \,
\int_{0}^{t} \frac{(\lambda t')^{2n}}{(2n)!}
\,e^{-\lambda t'}\,(\lambda dt')\,e^{-\lambda (t-t')} 
\,\phi_{-}(x)\nonumber \\
&= &e^{-\lambda t} \sinh \lambda t\,\phi_{-}(x)\,\,.
\label{Pminusplus}
\end{eqnarray}
While this is of course identical to $\phi_{-}(x)\,p(-,t\vert +)$ 
(and has no $x_{0}$-dependence), as 
one would expect, it is interesting to note that 
$P(x,-,t\vert x_{0}, -)$ is not simply equal to 
$\phi_{-}(x)\,p(-,t\vert -)$. Owing to the zero-transition contribution 
we obtain, instead,
\begin{equation}
P(x,-,t\vert x_{0}, -) = e^{-\lambda t} \left[ \delta (x-x_{0}) + 
\phi_{-}(x)\, (\cosh \lambda t - 1)\right]\,.
\label{Pminusminus}    
\end{equation}
The non-trivial cases (corresponding to a final state $\xi (t) =
+1$) can be analyzed similarly, to obtain the 
following results:
\begin{eqnarray}
\hspace{-0.7cm}P(x,+,t\vert x_{0},+)&=& e^{-\lambda t}\,\bigg[\delta 
\left(x- X(x_{0},t)\right) \nonumber \\ 
&+&\hspace{-0.3cm}\lambda \int_{-\infty}^{\infty} dx'
\int_{0}^{t} dt'\,\,\phi_{-}(x')
\,\,\delta (x - X(x',t-t'))\, \sinh \lambda t' \bigg]\,,
\label{Pplusplus}
\end{eqnarray}
and 
\begin{eqnarray}
P(x,+,t\vert x_{0},-) = \lambda e^{-\lambda t}\,
\int_{-\infty}^{\infty} dx' \int_{0}^{t} dt'\,\,\phi_{-}(x')\,\,
\delta (x - X(x',t-t'))\,\cosh \lambda t'\,.
\label{Pplusminus}
\end{eqnarray}
Here $X(x',t-t') = q^{-1}(q(x') +t-t')$, in accordance with the definition 
in Eq. (\ref{X}). 
The $x_{0}$-dependence 
of $P(x,+,t\vert x_{0},-)$ is implicit in Eq. (\ref{Pplusminus}); it
arises from the `one-transition'  
contribution to $P(x,+,t\vert x_{0},-)$, and is given by 
$\lambda \,t\,e^{-\lambda t}\,\phi_{-}(x_{0})$.
It is easily verified that the expressions in 
Eqs. (\ref{Pminusplus})-(\ref{Pplusminus}) are correctly normalized, 
according to   
\begin{equation}
\int_{-\infty}^{\infty} P(x,\pm,t\vert x_{0},\pm)\,dx =  p(\pm,t\vert 
\pm)\,\,, \,\,
\int_{-\infty}^{\infty} P(x,\pm,t\vert x_{0},\mp)\,dx
 =  p(\pm,t\vert \mp)\,\,.    
\label{Pnorm}      
\end{equation}
The integration over $x'$ in Eqs. (\ref{Pplusplus}) 
and (\ref{Pplusminus}) is obviously constrained by the 
$\delta$-function in each integrand: for a given $x$, it is 
restricted to those points lying on the trajectory $X(x',t-t')$
that are carried by the flow of Eq. (\ref{fplus}),
in a time ranging from $0$ to $t$, into the point $x$ at precisely 
the instant $t$.
Using 
the  $\delta$-functions to carry out the integration over $t'$ rather than 
$x'$ helps us express the results in a suggestive form. 
Equation (\ref{deltaXT}) implies that 
\begin{equation}
\delta \left(x - X(x',t-t')\right) =  
\frac{\delta \left(t'-t +T(x', x)\right)}{ \vert f_{+}(x) \vert}\,\,.    
\label{deltaXT1}      
\end{equation}    
Using this in Eqs. (\ref{Pplusplus}) and (\ref{Pplusminus}), we 
finally get   
\begin{eqnarray}    
P(x,+,t\vert x_{0},+)&=& e^{-\lambda t}\,\bigg[\delta 
\left(x- X(x_{0},t)\right) \nonumber \\ 
&+&\frac{\lambda}{f_{+}(x)} \int_{X(x,-t)}^{x} dx' \,\phi_{-}(x')\,
\sinh \lambda (t - T(x',x))  \bigg]\,\,,
\label{P++}
\end{eqnarray}
and 
\begin{eqnarray}
P(x,+,t\vert x_{0},-) = \frac{\lambda e^{-\lambda t}}{f_{+}(x)}
\int_{X(x,-t)}^{x} dx'\,\phi_{-}(x')\,\cosh \lambda (t - T(x',x)) \,\,.
\label{P+-}
\end{eqnarray}
We note that it is $f_{+}$, rather than $\vert f_{+} \vert$, that 
appears
in these final expressions: The limits of integration as they stand 
here automatically 
ensure the positivity of the probability densities concerned,
because $X(x,-t)$ lies to the left 
[right] of  $x$ if 
$f_{+}$ is positive [negative] in the relevant interval, for 
$t > 0$. 

\subsection{Stationary distribution in the active state}

It is trivially seen from Eqs. (\ref{Pminusplus}) and 
(\ref{Pminusminus}) that $\lim_{t \rightarrow 
\infty} P(x,-,t\vert x_{0}, \xi_{0}) = 
\frac{1}{2}\,\phi_{-}(x)$ as required. 
The interesting question is whether there exists a 
non-degenerate 
stationary distribution $\phi_{+}(x)$ in the {\it active} state, 
in accordance with Eq. (\ref{Ppluslimit}). In other words, what is
the eventual 
outcome of the competition between the attracting fixed point(s) 
towards which $x$ evolves each time it is in the active state, and 
the random re-setting it undergoes whenever it falls into the 
quiescent state? 

The more familiar approach to this question would be to start 
with the differential 
equation for $P(x,+,t\vert x_{0}, \xi_{0})$, namely, $(\partial_{t} +   
\partial_{x}\,f_{+} + \lambda )\,P_{+}(x,t) = \lambda \,P_{-}(x,t)$, in 
the limit $t \rightarrow 
\infty$: the stationary distribution $\phi_{+}(x)$ must 
satisfy the first 
order equation  
\begin{equation}
\left(\frac{d}{dx}f_{+}(x) + \lambda\right)\phi_{+}(x) = 
\phi_{-}(x)\,.      
\label{dephi+}      
\end{equation}
Moreover, it must be non-negative and normalizable. Equation (\ref{dephi+})
does make it clear that $\phi_{-}(x)$ acts as the 
`source' for $\phi_{+}(x)$. However, 
the disadvantage of 
this approach is that a constant of integration is explicitly 
involved: if $\phi_{+}(x) =
\bar{\phi}_{+}$ at $x=\bar{x}$, then the 
formal solution of Eq. (\ref{dephi+}) reads
\begin{equation}
\phi_{+}(x) = 
\bar{\phi}_{+} \,e^{-\int_{\bar{x}}^{x}F(x')dx'} + 
\int_{\bar{x}}^{x} dx'
\left(\frac{\phi_{-}(x')}{f_{+}(x')}\right)
e^{-\int_{x'}^{x}F(x'')dx''}
\label{phi+soln}      
\end{equation}
where $F(x) = (f_{+}'(x) +\lambda)/f_{+}(x)$. It is not immediately 
clear how to choose $\bar{x}$ and determine $\bar{\phi}_{+}$, and how 
the explicit dependence on these 
quantities disappears in
the final result for $\phi_{+}(x)$, as it must. Presumably, the other 
requirements on $\phi_{+}(x)$ 
such as its normalizability play a role here. As this is an 
interesting (and instructive) question in its own right, we shall 
return to it shortly. 
Before doing so, however, we point out that the 
exact time-dependent solution in Eq. (\ref{P++}) (or Eq. (\ref{P+-})) 
already provides us with the following complete answer, 
circumventing these problems.

Let $\alpha$, $\beta$ 
denote two successive zeroes of 
$f_{+}(x)$, and let the flow $\dot{x} = 
f_{+}(x)$ be directed from $\alpha$ to $\beta$ in the interval
($\alpha$, $\beta$). As $\alpha$ is a repellor, 
we have $X(x,-\infty) = \alpha$ for 
$x \in (\alpha, \beta)$. Therefore, passing to the limit 
$t \rightarrow 
\infty$ in Eq. (\ref{P++}) or (\ref{P+-}), we obtain
\begin{equation}    
\phi_{+}(x) = 
\frac{\lambda}{f_{+}(x)} \int_{\alpha}^{x} dx' \,\phi_{-}(x')\,
e^{-\lambda T(x',x)}\,\,,\,\, x \in (\alpha, \beta)\,.
\label{finalphi+}
\end{equation}
The same expression is valid in each such interval between successive 
critical points; the integration over $x'$ runs
{\it from the  
repellor} in that interval {\it to} the
point $x$. Equation (\ref{finalphi+}) not 
only specifies the exact solution for the stationary distribution in 
the active state, but also helps determine when such a solution 
exists, whether it is normalizable or not, when  a
pile-up or a divergence of the density occurs at the 
attracting fixed point, and when there is a ``leakage of probability"
into the absorbing state. As all these aspects are best illustrated by 
considering a set of typical examples as case studies, we now turn to
these. In what follows, $\gamma$ is a positive constant denoting a 
characteristic time scale of the deterministic dynamics given by 
$\dot{x} = f_{+}(x)$. \\
(i) {\it Stable drift}: Consider the simple case $f_{+}(x) = -\,\gamma 
x$, corresponding to a simple 
attractor at $x=0$. For definiteness, let us confine our attention to 
the region $x \geq 0$. The repellor in this region is $x = +\infty$, 
so that the formula  of Eq. (\ref{finalphi+})  
yields in this case the exact solution 
\begin{equation}    
\phi_{+}(x) = 
\frac{\lambda x^{-1+\lambda / \gamma}}{\gamma}\int_{x}^{\infty} 
\frac{\phi_{-}(x')}{(x')^{\lambda / \gamma}}\,dx' \,\,.
\label{phi+i}
\end{equation}
On the other hand, the formal solution given by  Eq. (\ref{phi+soln})
yields
\begin{equation}    
\phi_{+}(x) = 
x^{-1+\lambda / \gamma}\left(\frac{\bar{\phi}_{+}}
{\bar{x}^{-1+\lambda / \gamma}} - 
\frac{\lambda}{ \gamma}
\int_{\bar{x}}^{x} 
\frac{\phi_{-}(x')}{(x')^{\lambda / \gamma}}\,dx'\right)\,\,.
 \label{phi+i1}
\end{equation}
We require that  $\phi_{+}(x)$  be normalizable (i.e., integrable). 
However, the factor 
$x^{-1+\lambda / \gamma}$ leads to  a divergence 
at $+\infty$ (the repellor), unless the factor in parentheses 
in Eq. (\ref{phi+i1}) 
vanishes as $x 
\rightarrow +\infty$. Therefore we must impose the condition 
\begin{equation} 
\frac{\bar{\phi}_{+}}
{\bar{x}^{-1+\lambda / \gamma}} = 
\frac{\lambda}{ \gamma}
\int_{\bar{x}}^{\infty} 
\frac{\phi_{-}(x')}{(x')^{\lambda / \gamma}}\,dx'\,\,,
\label{phi+icondn}
\end{equation}
which not only removes the divergence, but also eliminates 
all dependence on $\bar{x}$ and $\bar{\phi}_{+}$ 
and leads to precisely the normalizable density given by Eq. 
(\ref{phi+i}).\\
(ii) {\it Unstable drift; power law tail} : Now let $f_{+}(x) = +\,\gamma x$, 
corresponding to an {\it unstable} critical point at the origin. We 
consider the region $x \geq 0$ as before. As $x=0$ is now the 
repellor, Eq. (\ref{finalphi+})  yields
\begin{equation}    
\phi_{+}(x) = 
\frac{\lambda}{\gamma \,x^{1+\lambda / \gamma}}\int_{0}^{x} 
\phi_{-}(x')\,(x')^{\lambda / \gamma}\,dx' \,\,.
\label{phi+ii}
\end{equation}
On the other hand, applying Eq. (\ref{phi+soln}) we obtain
\begin{equation}    
\phi_{+}(x) = 
\frac{1}{x^{1+\lambda / \gamma}}\left(\bar{\phi}_{+}
\,\bar{x}^{1+\lambda / \gamma} + 
\frac{\lambda}{ \gamma}
\int_{\bar{x}}^{x} 
\phi_{-}(x')\,(x')^{\lambda / \gamma}\,dx'\right)\,\,.
\label{phi+ii1}
\end{equation}
The problem of normalizability now arises from 
the factor $x^{-1-\lambda / 
\gamma}$, and again occurs at the repelling critical 
point (here, at $x=0$). To eliminate it, the factor in parentheses in Eq. 
(\ref{phi+ii1}) must be required to vanish as $x \rightarrow 0$. 
The condition to be imposed is therefore 
\begin{equation}  
\bar{\phi}_{+}\,
\bar{x}^{1+\lambda / \gamma} = 
\frac{\lambda}{ \gamma}
\int_{0}^{\bar{x}} 
\phi_{-}(x')\,(x')^{\lambda / \gamma}\,dx'\,\,,
\label{phi+iicondn}
\end{equation}
leading once again to the correct solution as given by Eq. (\ref{phi+ii}).

The following point is noteworthy. An inspection of Eq. (\ref{phi+ii}) shows
clearly that, in this case, the 
density $\phi_{+}(x)$ {\it always} has a power law tail.
(When $\phi_{-}(x)$ is an even function of $x$, 
replacing $x$ with $\vert x \vert $ in Eq. (\ref{phi+ii}) yields 
$\phi_{+}(x)$  in the entire range $-\infty < x < \infty$.) 
The 
random re-setting of $x$ in the quiescent state does compensate for 
the unstable 
drift towards infinity in the evolving state, but only to the extent of 
producing a 
normalizable stationary distribution in the latter state. It is
insufficient to prevent a 
slow (power-law) fall-off of the latter, no matter how rapidly 
the fixed density
$\phi_{-}(x)$ falls off for large $\vert x \vert $, or even if its 
support is compact. Even then, the variance of $x$ diverges unless 
$\lambda / \gamma > 2$, i.e., 
the switching rate is greater than twice the drift coefficient.

This example also represents a very direct 
way in which {\it simple dynamics can generate the whole family of stable 
(Levy)
distributions.} As the (cumulative) distribution function of $x$ has a 
tail $\sim x^{-\lambda / \gamma}$, a (suitably re-scaled 
and shifted) sum of such variates, independently distributed, will have a 
stable distribution with exponent $\lambda / \gamma$, going over into 
a normal distribution for $\lambda / \gamma \geq 2$.
\\  
(iii) {\it Higher order critical point; `leakage' of probability} : 
Next, consider the case $f_{+}(x) = \gamma x^{2}$.
The flow is towards $x=0$ for all negative $x$, 
while it is directed away from this point and towards $+\infty$ for 
positive $x$.  As before, 
the exact solution for $\phi_{+}(x)$ may be found by using Eq. 
(\ref{finalphi+}), or by using Eq. $(\ref{phi+soln})$ and imposing the 
conditions required to eliminate divergences and ensure 
normalizability
(at $-\infty$ for negative $x$, 
and at $x=0$
for positive $x$). The result is, for $x>0$, 
\begin{equation}
\phi_{+}(x) = 
\frac{\lambda e^{\lambda / (\gamma x)}}{\gamma x^{2}}
\int_{0}^{x} \phi_{-}(x')\,
e^{-\lambda / (\gamma x')}\,dx'\,\,.
\label{phi+iii}
\end{equation}
For $x<0$, the lower limit of integration is
$-\infty$ instead of $0$.

As in the preceding case, $\phi_{+}(x)$ has a power law tail. But  
the degenerate critical point at the origin has an even stronger 
effect. Any positive initial value $x_{0}$ reaches  
$+\infty$ in a {\it finite} time $(\gamma x_{0})^{-1}$ under the flow 
in the evolving state. We may therefore expect some sort of  
``absorption" at 
$+\infty$: in terms of probability distributions, this would show up as 
a ``leakage of probability" leading to a deficit in the total 
probability. This is indeed borne out: we find that
\begin{equation}
\int_{-\infty}^{0}\phi_{+}(x)\,dx = 
\int_{-\infty}^{0}\phi_{-}(x)\,dx\,\,, 
\label{phi+iiinormleft}
\end{equation}
but
\begin{equation}
\int_{0}^{\infty}\phi_{+}(x)\,dx = 
\int_{0}^{\infty}\phi_{-}(x)\,(1- e^{-\lambda / (\gamma x)})\,dx\,\,. 
\label{phi+iiinormright}
\end{equation}
This loss in probability measure can also be given a simple 
interpretation. 
The probability of being injected into $(x, x+dx)$ under the random 
re-setting is $\phi_{-}(x)dx$. The time taken to reach $+\infty$ 
from any $x > 0$ under the flow is $(\gamma x)^{-1}$. The 
probability of remaining in the evolving state for this duration is 
$\exp (-\lambda / \gamma x)$. Therefore the total loss in measure is  
$\int_{0}^{\infty}\phi_{-}(x) \exp (-\lambda / \gamma x) dx$. In 
general, such a loss of measure occurs whenever the absorbing state 
at the attracting critical point 
is reached in a finite time under the deterministic flow, as will 
become clear from the examples 
to follow.

It is also an instructive exercise to ``unfold" the degenerate 
critical point at $x=0$ by 
starting, for instance,  with $f_{+}(x) =\gamma 
x(x-\epsilon)$, and then examining how various quantities behave as 
$\epsilon \rightarrow 0$.\\ 
(iv) {\it Periodic boundary conditions; single critical point} : In 
many physical applications, $x$ is an angular variable, so that its 
range is compact, and
periodic boundary conditions apply. Consider the case $\dot{x} = 
\gamma\,\vert \sin x \vert$, with a fundamental period equal to $\pi$. 
There is only one critical point, and the flow is directed from $0$
towards $\pi$. Given that $\phi_{-}(x) = \phi_{-}(x+ \pi)$, we seek a 
solution $\phi_{+}(x)$ that has the same periodicity property. Using   
Eq. (\ref{finalphi+}), we find
\begin{equation}    
\phi_{+}(x) = 
\frac{\lambda}{ \gamma \sin x}\int_{0}^{x} 
\left(\frac{\tan x'/2}{\tan x/2}\right)^{\lambda / \gamma}\,
\phi_{-}(x')\,dx' \,\,,\,\, x \in [0, \pi)\,,
\label{phi+iv}
\end{equation}
together with $\phi_{+}(x) = \phi_{+}(x+ \pi)$. The time taken to 
reach $\pi$ from any $x$ in $0 < x < \pi$ is infinite (although the 
range is finite, the velocity $f_{+}(x)$ vanishes as $x \rightarrow 
\pi$). Therefore there is no loss of measure in this case. However, a 
pile-up of the density can occur at $x = \pi$, depending 
on the value of the ratio
$ \lambda/ \gamma $: for a uniform $\phi_{-}(x)$, for instance, we 
can show from Eq. (\ref{phi+iv}) that $\phi_{+}(x)$ is finite at $x = \pi$
only if the switching rate $ \lambda > \gamma $. As $x \uparrow 
\pi$, we find $\phi_{+}(x) \sim - \ln \, (\pi - x)$ for $ \lambda= 
\gamma $, and $\phi_{+}(x) \sim (\pi - x)^{-1 +\lambda/ \gamma}$ for 
$ \lambda < \gamma $. Thus $\phi_{+}(x)$ is divergent (though 
integrable) at the attractor, unless the switching rate exceeds 
the drift coefficient.\\
(v) {\it Periodic boundary conditions; degenerate critical point} : 
As the periodic analog of Case (iii) above, consider $f_{+}(x) = 
\gamma\,\sin^{2} x $ with the fundamental interval $[0, \pi)$. We find 
the periodic solution  
\begin{equation}    
\phi_{+}(x) = 
\frac{\lambda}{ \gamma \sin^{2} x}
e^{(\lambda / \gamma) \cot x}
\int_{0}^{x} 
\phi_{-}(x')\,e^{-(\lambda / \gamma) \cot x'}\,
dx' \,\,\,,\,\, x \in [0, \pi)\,\,.
\label{phi+v}
\end{equation}
Again, as the time taken to 
reach $\pi$ from any $x$ in $0 < x < \pi$ is infinite, there is no 
loss of measure owing to absorption at $\pi$. Moreover, now there is 
no
divergence of $\phi_{+}(x)$ at $x = \pi$, either, because the velocity 
$f_{+}(x)$ vanishes more rapidly than linearly as $x \rightarrow 
\pi$.

A flow like $\dot{x} = 
\gamma \, x$  with periodic boundary conditions ([0, 1] being the 
fundamental interval, for instance), {\it will} produce a leakage of
probability into the (finite) absorbing point at $x=1$:
now the velocity does not vanish as  
$x \uparrow 1$. It is easily shown that 
$\int_{0}^{1}\phi_{+}(x)\,dx = 
\int_{0}^{1}\phi_{-}(x)\,(1- x^{\lambda / \gamma })\,dx$ in this 
case.\\
(vi) {\it Periodic boundary conditions; two critical points} : 
Finally, let us consider the case $\dot{x} = 
\gamma\, \sin x $, with a period equal to $2\pi$. The velocity 
$f_{+}(x)$ changes sign in the fundamental interval, and  
there is a repellor (at $0$) as well as an attractor (at $\pi$) in it. 
In the range $0 \leq x <\pi$, the 
solution for $\phi_{+}(x)$ is exactly as in Eq. (\ref{phi+iv});
for $\pi \leq x <2\pi$, the lower limit of integration is 
replaced by $2\pi$. The divergent behavior found as $x \uparrow 
\pi$ when $ \lambda \leq \gamma $ in Case (iv) above now occurs 
symmetrically on both sides of $x = \pi$, as one would expect. 

\section{Stationary distribution in dichotomous flow: some remarks}

The form of the stationary distribution in Eq. (\ref{finalphi+})
and the discussion following it suggest a direct physical 
interpretation (amounting, in fact,  to a simple heuristic derivation) 
of the well-known formula \cite{kitahara}, \cite{horst'} 
for the stationary distribution that obtains (under certain 
conditions) in
the case of the full-fledged dichotomous flow given by the stochastic
differential equation
\begin{equation}
\dot{x} = f(x) +g(x) \xi (t)\,\,.
\label{sde}
\end{equation}
Such a flow corresponds to the random alternation of two 
deterministic flows, given by $\dot{x} = f_{+}(x)$ and $\dot{x} 
= f_{-}(x)$, respectively, where $f_{\pm}(x) \equiv f(x) \pm g(x)$. 
It serves as a model for numerous physical phenomena (see, e.g., 
Refs. \cite{horst'}, \cite{vdb}).

The time-dependent probability densities 
$P(x,\xi,t\vert x_{0},\xi_{0})$ (denoted by $P_{\pm}(x,t)$ for short)
now satisfy the coupled first order equations
$(\partial_{t} +   
\partial_{x}\,f_{\pm} + \lambda )\,P_{\pm}(x,t) = \lambda 
\,P_{\mp}(x,t)$ . In general, it is not possible to obtain
a partial differential  equation of {\it finite} order for 
the total probability density $P(x,t) \equiv P_{+}(x,t) + P_{-}(x,t)$:
this possibility is restricted to certain special cases 
\cite{jung}-\cite{bb}.
However, if the deterministic flow is ``dynamically stable", 
the {\it stationary} density $P^{st}(x) = \lim_{t \rightarrow 
\infty} P(x,t)$ is found to satisfy the first order ordinary 
differential equation
\begin{equation}
\frac{d}{dx}\left((f+g)(1-\frac{f}{g})P^{st}\right)
-2\lambda \frac{f}{g} P^{st} = 0\,\,. 
\label{Pstateq}
\end{equation}
Here, the 
term  ``dynamic 
stability" denotes precisely the situation of interest in the 
present context: the existence of two different stable critical points 
in the flows   
$\dot{x} = f_{+}(x)$  and  $\dot{x} 
= f_{-}(x)$ respectively, with no other critical point of either flow 
in between them.  What is sought is the stationary 
distribution to which $x$ settles down under the competition between 
the two attractors. (Thus the second attractor takes on the role 
played in the preceding section by the 
random re-setting of $x$.) A typical case that helps visualize 
the situation is as follows: a repellor at $x = \beta$ and 
an attractor at $x = b$ in the flow 
$f_{+}(x)$; and an attractor at $x = a$ and a repellor at $x = \alpha$
in the flow $f_{-}(x)$. If $\beta \leq a < b \leq \alpha$, the 
interval  $[a,b]$ acts as a trapping region 
supporting a non-trivial  
$P^{st}(x)$. The latter is given \cite{kitahara},\cite{horst'} by 
the solution of Eq. (\ref{Pstateq}), namely,
\begin{equation} 
P^{st}(x) = {\mathcal N} \frac{g}{g^{2}-f^{2}} \exp \left(2\lambda \int
\, \frac{f\,dx}{g^{2}-f^{2}}\right)\,\,,
\label{Pstat}
\end{equation}
where $\mathcal N$ is the normalization constant. We 
give a simple interpretation of this formula along the lines of that 
presented for Eq. (\ref{finalphi+}) above, which suggests how the
expression in Eq. (\ref{Pstat}) can virtually be written down by
inspection. 

Consider, first, what happens if the system alternates
{\it regularly } between 
the two flows, with a constant duration $\tau$ in each state. Let us 
denote 
the solution trajectories in the two flows by $X_{+}(x,t)$ and 
$X_{-}(x,t)$, respectively. Thus, if we start in the $+$ state 
from the point $x_{0}$, we have $x(\tau) = X_{+}(x_{0},\tau)$,
$x(2\tau) = X_{-}(x(\tau),\tau)$, and so on. Setting $x(2n\tau) 
\equiv \eta_{n}$ and $x((2n+1)\tau) 
\equiv \zeta_{n}$ (where $n = 0,1,\ldots$), the dynamics is given by 
the two-dimensional map 
\begin{equation} 
\zeta_{n} = X_{+}(\eta_{n})\,\,, \,\,\eta_{n} = X_{-}(\zeta_{n})\,\,.
\label{2Dmap}
\end{equation}
In the ``dynamically stable" situation referred to above, there exist 
stable fixed points $\eta^{*}$ and $\zeta^{*}$ such that $\eta^{*} 
= X_{-}(X_{+}(\eta^{*}))$, $\zeta^{*}  = X_{+}(X_{-}(\zeta^{*}))$, implying 
a stable period-two cycle into which the original variable $x(n\tau)$  
falls asymptotically as $n \rightarrow \infty$. Correspondingly, 
in continuous time the trajectory  $x(t)$ zig-zags between the values 
$\eta^{*}$ and $\zeta^{*}$, 
alternately following the two evolution rules. An invariant density 
for $x$ can therefore be defined in each of the two distinct 
alternating segments. Calling these $\phi_{+}(x)$ and $\phi_{-}(x)$, 
respectively, it is evident that $\phi_{\pm}(x) \vert dx \vert_{\pm} = 
\tau^{-1} dt$, or $\phi_{\pm}(x) = (\tau\,\vert f_{\pm}(x) 
\vert)^{-1}$, i.e., essentially just the reciprocal of 
the corresponding Jacobian. 
As the $+$ and $-$ states are equally probable, the invariant density 
of $x$ becomes 
\begin{equation} 
P^{st}(x) = \frac{1}{2\,\tau}\left(\frac{1}{\vert f_{+}(x) \vert} +
\frac{1}{\vert f_{-}(x) \vert}\right)\,\,.
\label{Pstregular}
\end{equation}
Taking into account the fact that the flows are in opposite directions,
i.e., that $f_{+}(x)$ and  $f_{-}(x)$ have 
opposite signs, the quantity in parentheses in Eq. 
(\ref{Pstregular}) is proportional to $f_{-}^{-1} - f_{+}^{-1} =
g(g^{2}-f^{2})^{-1}$. This 
explains the origin of the factor preceding the exponential in
Eq. (\ref{Pstat}).  

It remains to understand the exponential factor in Eq. (\ref{Pstat}).
Returning to the case of random switching between the two flows, 
it is evident that the weight factor $\tau^{-1}$ in Eq. 
(\ref{Pstregular}), applicable when the switching occurs at 
regular
intervals, must be replaced by  a product of two probability factors,
one for each of the two flows. This factor is 
the same as that already discussed in relation with Eq. 
(\ref{finalphi+}) in the case of interrupted evolution, namely, an
exponential of the time required to reach the point $x$ from the 
corresponding repellor - or rather, from the attractor
in the {\it other} flow, recalling 
that the trapping region (the support of $P^{st}(x)$)
is bounded by these points. Thus, in the 
example mentioned above (in which $\beta \leq a < b \leq \alpha$)), 
we have the factor 
$\exp [- \lambda T_{+}(a,x)] \exp [- \lambda T_{-}(b,x)]$, where 
we have used subscripts on $T$ to indicate the corresponding flow. 
Since $T_{\pm}(u,x) = \int_{u}^{x} dx'/f_{\pm}(x')$ and $f_{\pm} = f \pm 
g$, we find
\begin{eqnarray}
T_{+}(a,x) + T_{-}(b,x) &=& \int_{a}^{x} \frac{dx'}{g(x') + f(x')} 
+\int_{x}^{b} \frac{dx'}{g(x') - f(x')}\nonumber \\
&=& -2 \int_{a}^{x} \frac{f\,dx'}{g^{2} - f^{2}} + \mbox{const.}
\label{T+T}
\end{eqnarray}
This yields precisely the exponential factor occurring in 
the solution given by Eq. (\ref{Pstat}).  
The ``mechanism" underlying this formula is therefore made more 
manifest by re-writing it in the form  
\begin{equation} 
P^{st}(x) = 
\frac{{\mathcal N} }{2}\left(\vert f_{+}(x) \vert^{-1} +
\vert f_{-}(x) \vert^{-1}\right)\,
e^{- \lambda\,T_{+}(a\,,\,x)}\,e^{-\lambda\,T_{-}(b\,,\,x)}
\label{Pstatnew}
\end{equation}
for the configuration of attractors under consideration, with 
obvious minor modifications for the other possible configurations.

A simple example is provided 
by the exactly solvable case \cite{sancho} of linear dichotomous flow,   
$\dot{x} = -\gamma \,x + c \,\xi (t)$, where $\gamma$ and $c$ are positive 
constants. Here $\beta = - \infty, \,a = -c/ \gamma,\, b = c/ \gamma, 
\,\alpha = \infty$. The stationary density
of $x$ has support in the interval
$[-c/ \gamma\,,\, c/ \gamma]$, and is proportional to $(c^{2} - 
\gamma^{2}x^{2})^{-1 + \lambda / \gamma}$. We can now readily understand
its interesting features as follows. Since $T_{+}(x_{0},c/ \gamma)$ 
and $T_{-}(x_{0}, -c/ \gamma)$ diverge, there is no loss of 
probability measure owing to the absorption at the attractors 
in this instance. 
However, as we may now anticipate (based on the remarks made 
in Case (iv) above), $P^{st}(x)$ is indeed
divergent at the (finite) endpoints $\pm \,c/ \gamma$ when the 
switching rate  $\lambda$ is smaller than the 
drift coefficient $\gamma$.   

\section*{Acknowledgments}

This work was supported in part by the Interuniversity Attraction Poles 
Program of the Belgian Federal Government. VB acknowledges the warm 
hospitality of the Limburgs Universitair Centrum and the Universit\'e 
Libre de Bruxelles.


\begin{thebibliography}{99}
\bibitem{poisson} This is the assumption generally made in modeling
the effective Liouville operator corresponding to stochastically 
interrupted deterministic evolution  (or unitary evolution, in
the quantum mechanical case) in most physical applications 
(see, e.g., Ref. \cite{clauser}). It represents a kind of Markovian 
assumption (e.g.,  $x$ then turns out to be a component of a Markov
process). Non-Markovian generalizations based on interruptions
governed by an arbitrary  renewal process rather than a 
Poisson  process are possible \cite{bala}.
However, as stated earlier, our present emphasis is on a
different aspect of the problem.
\end{thebibliography}
\end{document}